\documentclass[twocolumn,showpacs,preprintnumbers,amsmath,amssymb]{revtex4}

\usepackage{amsmath}
\usepackage{graphicx}
\usepackage{calc}
\usepackage{bm}
\usepackage{float}

\begin{document}

\title{Spin polarization and exchange-correlation effects in transport properties of two-dimensional electron systems in silicon}
\author{V. T. Dolgopolov}
\affiliation{Institute of Solid State Physics, Chernogolovka, Moscow District 142432, Russia}
\author{A. A. Shashkin}
\affiliation{Institute of Solid State Physics, Chernogolovka, Moscow District 142432, Russia}
\author{S. V. Kravchenko}
\affiliation{Institute of Fundamental and Frontier Sciences, University of Electronic Science and Technology of China, Chengdu 610054, People's Republic of China}
\affiliation{Physics Department, Northeastern University, Boston, Massachusetts 02115, USA}

\begin{abstract}
We show that the parallel magnetic field-induced increase in the critical electron density for the Anderson transition in a strongly interacting two-dimensional electron system is caused by the effects of exchange and correlations. If the transition occurs when electron spins are only partially polarized, additional increase in the magnetic field is necessary to achieve the full spin polarization in the insulating state due to the exchange effects.
\end{abstract}
\maketitle

The metal-insulator transition (MIT) in two-dimensional (2D) electron systems, studied experimentally and theoretically in 1970s \cite{ando}, was declared nonexistent after negative logarithmic quantum corrections to the conductivity had been found (for a review, see Ref.~\cite{lee1985}). The reasoning was as follows. In an infinite 2D system, upon decreasing temperature, negative quantum logarithmic corrections to the conductivity will eventually become comparable to the conductivity itself. After this, conductivity will decrease exponentially. Therefore, the system will inevitably become an insulator no matter how high the initial value of the conductivity is. However, it has later been shown both theoretically \cite{fink1,lee,fink2,fink} and experimentally \cite{kra1,shashkin2005} that this conclusion may be wrong in 2D systems with strong electron-electron interactions.

Since there has been a certain amount of confusion and controversy in the literature regarding the zero-temperature MIT in infinite 2D systems (see, \textit{e.g.}, Ref.~\cite{gant}), here we will consider a disorder-driven Anderson MIT at finite (although low) temperatures and in finite 2D systems. (As correctly stated in Ref.~\cite{Wain1}, the question about the true MIT is ``a rather academic question as what has actually been measured experimentally corresponds to rather high energy physics''.) Attempts to describe the experimentally observed behavior of the critical density for the MIT in silicon metal-oxide-semiconductor field-effect transistors (MOSFETs) as a function of a parallel to the interface magnetic field, $B$, were made by quite a few theoretical groups \cite{Wain1,fl,Sarm1,Sarm2,Dobr,am}. Nevertheless, the satisfactory explanation of experimental results is still absent. Monte Carlo calculations and finite size scaling techniques \cite{Wain1,fl} show that the spin polarization in strongly-correlated electron systems favors localization. Using the appraisal Ioffe-Regel criterion to calculate the critical density in the Born approximation with two fitting parameters, the authors of Ref.~\cite{Sarm1} have achieved a satisfactory agreement with the experiment, but correlation effects have not been taken into account and their negligibility in the case of strong electron-electron interactions has not been established. Doubts in the applicability of the percolation scenario to the transition in Si MOSFETs \cite{Sarm2} are expressed in Ref.~\cite{Sarm1}. The increase of the effective mass with decreasing electron density suggests that the effect of interactions is the dominant driving force for the experimentally observed MIT due to fermion condensation \cite{am} or the Wigner-Mott transition \cite{Dobr,dov}. Indeed, theory \cite{Dobr} is in excellent qualitative agreement with the experiment. However, the critical electron densities measured in the experiment are sample-dependent even for similar samples \cite{brun} pointing to the importance of the disorder. This forces us to reconsider a disorder-driven Anderson MIT.

In this paper, we compare the experimental data on electron transport in a strongly correlated 2D electron system in (100) silicon MOSFETs with maximum mobility of $\sim3\times10^4$~cm$^2$/Vs subjected to a parallel magnetic field with the results of the calculations described below. We also compare the data for the critical density of the MIT as a function of the magnetic field with the dependence of the complete spin polarization field $B^\text{p}$ on electron density, $n_\text{s}$. We show that within the theory of disorder-driven Anderson MIT, the localization of unpolarized electrons in zero magnetic field occurs at a lower electron density, $n_\text{c}$, than that of fully spin-polarized ones at density, $n_\text{c1}$, which is caused by the exchange and correlation effects. To check the relevance of the theory independently, we calculate the resistance ratio of the spin polarized and spin unpolarized electron systems in the metallic regime near the MIT. Using the impurity density, $N_\text{i}$, as the only adjustable parameter, we describe the experimental results for both the resistance and $n_\text{c1}/n_\text{c}$ ratios. Although the data similar to those used in this paper have been obtained by many experimental groups (see, \textit{e.g.}, Refs.~\cite{pu,vit,drag}), for a detailed comparison, data obtained on the same sample are needed; therefore, we only use our own data in the analysis.

To experimentally determine the position of the critical density for the Anderson transition, we have used two methods. (i) In the insulating state, the conductivity measured at currents approaching zero has an activated character. The transition point corresponds to the electron density at which the activation energy $\Delta=0$. (ii) Current-voltage characteristics in the insulating state are strongly nonlinear. This nonlinearity vanishes at the transition point. (Similar criteria were used in Ref.~\cite{diorio92} to determine the phase diagram of the re-entrant insulating phase in Si MOSFETs in perpendicular magnetic fields.) Both methods yield identical results \cite{shashkin2001}.

\begin{figure}
\scalebox{0.45}{\includegraphics[angle=0]{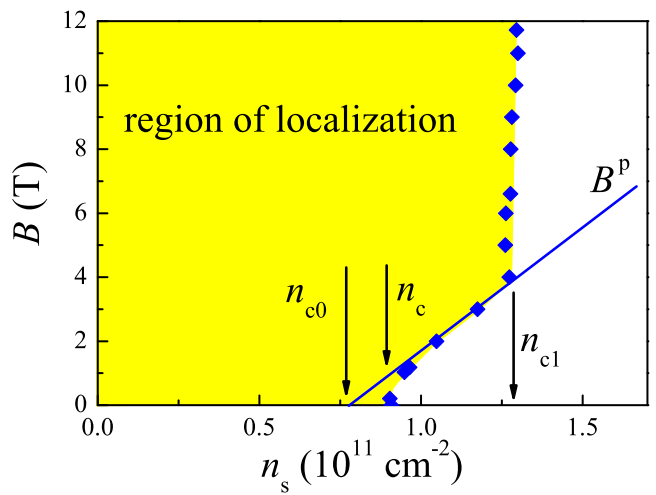}}
\caption{\label{fig1} Experimentally determined critical densities for the Anderson transition (diamonds) \cite{sh}. The solid line corresponds to the onset of the full spin polarization \cite{sh1}.}
\end{figure}

\begin{figure}\vspace{.1cm}
\hspace{0.1 cm}
\scalebox{0.43}{\includegraphics[angle=0]{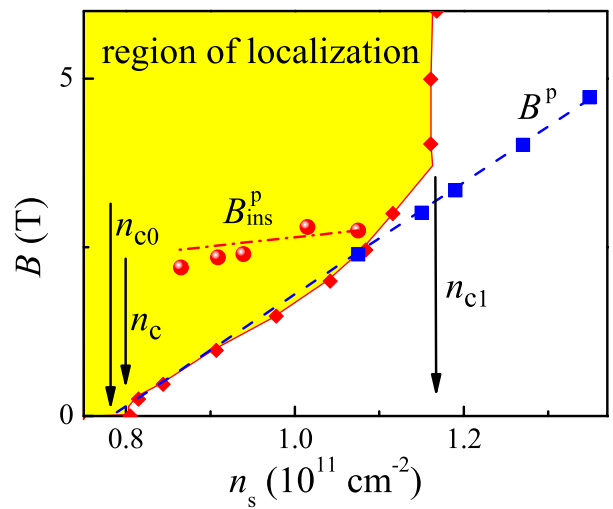}}
\caption{\label{fig2} Experimentally determined boundary for Anderson localization (diamonds) \cite{shashkin2001} and for the onset of the full spin polarization (squares) \cite{sh1}. Circles denote the field of the complete spin polarization in the insulator, taken from Ref.~\cite{met}. Dot-dashed line is the result of calculations using Eq.~(\ref{8}) with $\beta = 0.06$.}
\end{figure}

The experimental data are shown for two samples from Ref.~\cite{sh} in Fig.~\ref{fig1} and Ref.~\cite{shashkin2001} in Fig.~\ref{fig2} where the critical densities for the Anderson transition in parallel magnetic fields are compared to the position of the onset of the full spin polarization. Both samples have the same maximum electron mobility, but the quality of the sample shown in Fig.~\ref{fig2} is higher (presumably due to a better homogeneity), which results in somewhat smaller critical electron densities for the Anderson transition.

As seen from the figures, there are three distinct electron densities on the $x$-axis: $n_\text{c0}$ obtained by extrapolating the magnetic field of the complete spin polarization to zero \cite{sh1}, $n_\text{c}$ corresponding to the Anderson transition in zero magnetic field, and $n_\text{c1}$ corresponding to the onset of saturation of the transition point with the magnetic field. Since the magnetic field parallel to the interface only aligns electron spins (diamagnetic effects in Si MOSFETs used in this work are small \cite{kra2,pud}), density $n_\text{c1}$ corresponds to the MIT in a fully spin-polarized electron system. All three characteristic electron densities lie in the regime of strong electron-electron interactions. Note that the strength of the Coulomb interactions is usually characterized by the parameter $r_\text{s}=(\pi n_\text{s})^{-1/2}/a^*$, where $a^*$ is the effective Bohr radius; strongly correlated regime is achieved when $r_\text{s}\gg1$.

We will not discuss the clean limit in the highest quality samples in which $n_\text{c}$ practically coincides with $n_\text{c0}$, suggesting that the zero-field MIT in such samples is driven by interactions \cite{kra1,shashkin2005}. The Anderson MIT to be considered here occurs in more disordered samples in which $n_\text{c}$ exceeds $n_\text{c0}$, pointing to the importance of the disorder.

To calculate critical electron densities $n_\text{c}$ and $n_\text{c1}$, we will assume that the localization of electrons in Si MOSFETs is the result of the electron scattering on the random potential with the density of impurities equal to $N_\text{i}$. As shown in numerous papers starting with Ref.~\cite{got}, the localization of strongly interacting electrons is the result of the multiple electron scattering affected by screening in the presence of the exchange-correlation effects. We will suppose that the impurities are distributed in the 2D layer with zero thickness and neglect the quantum corrections due to finite temperature of experiments \cite{remark}. The 2D electron system is presumed to be homogeneous, in contrast to Ref.~\cite{kunc}. The distance between the gate and the 2D electron system is supposed to be large, $d\gg n_\text{c}^{-1/2}$, so that the screening by gate is negligible. The authors of Refs.~\cite{got,Be,G1,G2,G4} used their results without restrictions, although formally one should expect them to be valid only at $r_\text{s}\lesssim1$. Below we show that our calculations yield a satisfactory agreement with the experiment even at $r_\text{s}\sim10$.

In the vicinity of the metal-insulator transition, the conductivity can be written \cite{Be,G1}
\begin{equation}
\sigma=\sigma_0(1-A),\label{1}
\end{equation}
where $\sigma_0$ is the conductivity calculated in the Born approximation and $A$ is given by Ref.~\cite{G2}

\begin{equation}
A =\frac{1}{4\pi n_\text{s}^2} \int_0^\infty dq q \frac{<|U(q)|^2> X_0(q)^2}{(1+V(q)[1-G(q)]X_0(q))^2}.\label{2}
\end{equation}
Here $X_0(q)$ is the Lindhard function (see, \textit{e.g.}, Ref.~\cite{ando}), $<|U(q)|^2> = N_\text{i}(2\pi e^2/\varepsilon q)^2$, $V(q)=2\pi e^2/\varepsilon q$, $\varepsilon = 7.7$ is the average dielectric constant, and $G(q)$ is the Hubbard correction. The critical electron density $n_\text{c}$ corresponds to
\begin{equation}
A(n_\text{c})=1.\label{3}
\end{equation}
Since in the immediate vicinity of $n_\text{c}$, the Fermi wavevector is not a good quantum number, the critical density determined by Eq.~(\ref{3}) corresponds to the value obtained by extrapolation of the conductivity to zero from the metallic region.

We are interested in the transitions in both unpolarized and spin-polarized electron systems. A standard expression for the Hubbard correction is

\begin{equation}
G(q)=\frac{1}{g_\text{s} g_\text{v}}\frac{q}{(q^2+k_\text{F}^2)^{1/2}},\label{4}
\end{equation}
where $g_\text{s}$ and $g_\text{v}$ are spin and valley degeneracy, correspondingly, and $k_\text{F}$ is the Fermi wavevector.

Equation (\ref{2}) can be rewritten in a more convenient form

\begin{equation}
A =\frac{N_\text{i}}{4\pi n_\text{s}^2}\int_0^\infty dq q\frac{ (q_\text{s}/q)^2}{(1+[1-G(q)]q_\text{s}/q)^2},\label{5}
\end{equation}
where $q_\text{s}=q_\text{s0}$ for $q\leq 2k_\text{F}$ and $q_\text{s}=q_\text{s0}[1-(1-(2k_\text{F}/q)^2)^{1/2}]$ for $q>2k_\text{F}$. Here $q_\text{s0}= g_\text{s}g_\text{v}m^\ast e^2/\varepsilon\hbar^2$ is the inverse screening radius and $m^\ast$ is the effective mass.

\begin{figure}\hspace{0.1 cm}
\scalebox{0.45}{\includegraphics[angle=0]{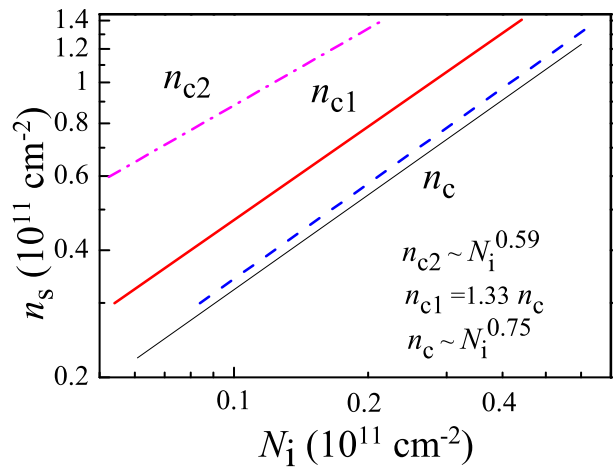}}
\caption{\label{fig3} Critical densities for the Anderson transition in a fully spin-polarized (solid line), spin-unpolarized (dashed line), and fully spin- and valley-polarized (dash-dotted line) electron systems as functions of the density of impurities. Thin solid line corresponds to $n_\text{c}\propto N_\text{i}^{0.75}$.}
\end{figure}

The integral in Eq.~(\ref{5}) can be easily evaluated in the limit $q_\text{s0}\gg 2k_\text{F}$. The main contribution comes from $q<q_0$; $2k_\text{F}\ll q_0< 2q_\text{s}(1-G(q_0))$;

\begin{equation}
q_0^3 \simeq 4q_\text{s0}k_\text{F}^2(1-G(\infty)).\label{6}
\end{equation}
Replacing the upper limit in Eq.~(\ref{5}) with $q_0$, one gets

\begin{equation}
A \simeq \frac{N_\text{i}}{2\pi n_\text{s}^{4/3}\left(1-G(\infty)\right)^{4/3}} \left[\frac{2\pi q_\text{s0}}{g_\text{s}g_\text{v}}\right]^{2/3}.\label{7}
\end{equation}

Although equation (\ref{7}) can hardly be applied to the experimentally studied Si MOSFETs, it is useful. First, it shows that the difference in critical electron densities in spin-polarized and spin-unpolarized systems is due only to the exchange-correlation effects that are described in our case by Hubbard corrections, and secondly, the critical density in a spin-polarized system turns out to be higher (approximately by a factor of 1.5) than that in a spin-unpolarized system.

As follows from Eqs.~(\ref{3}) and (\ref{5}), in the opposite limiting case $q_\text{s0}\ll k_\text{F}$ (\textit{i.e.}, when $r_\text{s}\ll 1$), one arrives at the opposite conclusion: $n_\text{c} >n_\text{c1}$ and $n_\text{c}<N_\text{i}$.

We have numerically solved Eqs.~(\ref{4}) and (\ref{5}) in the limit $q_\text{s0}\gg 2k_\text{F}$. The results for $n_\text{c}(N_\text{i})$ and $n_\text{c1}(N_\text{i})$ are shown in Fig.~\ref{fig3}. As seen in the figure, in quite a wide range of parameters, both critical densities are proportional to each other with the coefficient of proportionality of approximately 1.33. Their dependence on the impurity density obeys a power law: $n_\text{c}\propto n_\text{c1}\propto N_\text{i}^{0.75}$, which is in agreement with Eq.~(\ref{7}). As has been mentioned in Ref.~\cite{G4}, each impurity localizes more than one electron.

\begin{figure}\hspace{0.1 cm}
\scalebox{0.4}{\includegraphics[angle=0] {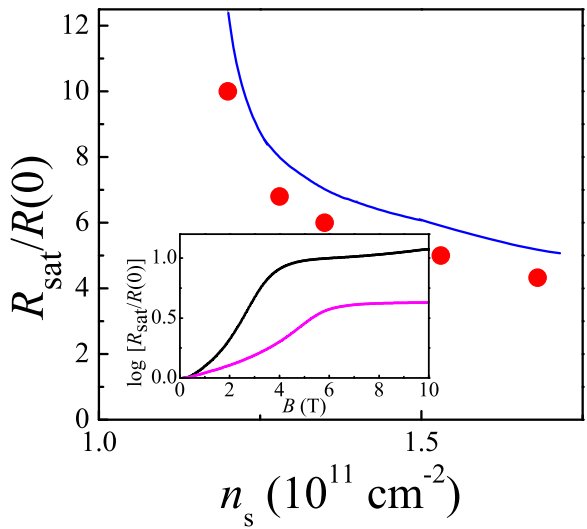}}
\caption{\label{fig4} The ratio of the resistances of spin-polarized and spin-unpolarized electrons at $T=30$~mK as a function of electron density. Circles correspond to the experimental data of Ref.~\cite{sh1} and the solid line is the result of calculations without additional fitting parameters. The inset shows the experimental curves at electron densities $1.2\times10^{11}$~cm$^{-2}$ (upper curve) and $1.68\times10^{11}$~cm$^{-2}$.}
\end{figure}

According to the data shown in Fig.~\ref{fig1}, $n_\text{c}$ is approximately equal to $9\times10^{10}$~cm$^{-2}$, which yields the density of impurities $\approx3.6\times10^{10}$~cm$^{-2}$ and $n_\text{c1}\approx1.2\times10^{11}$~cm$^{-2}$, both values agreeing reasonably well with the experiment. Moreover, the obtained density of impurities is in remarkably good agreement with that obtained independently from the results of Refs.~\cite{sh,d}.

Calculations using the data shown in Fig.~\ref{fig2} where $n_\text{c}\approx8\times10^{10}$~cm$^{-2}$ yield $N_\text{i}\approx3.1\times10^{10}$~cm$^{-2}$ and the expected $n_\text{c1}$ equal to $\approx1.07\times10^{11}$~cm$^{-2}$ that is in good agreement with the experimental value $n_\text{c1}\approx1.17\times10^{11}$~cm$^{-2}$.

It is easy to calculate the critical density $n_\text{c2}$ for the MIT in a fully spin- and valley-polarized electron system (Fig.~\ref{fig3}). However, comparison with the experiment is impossible because the available experimental data \cite{Ren} correspond to the insulating state.

In spite of the satisfactory agreement with the experiment, one should treat the results of the above calculations with some reservations. They will undoubtedly change if one uses other (more accurate) values of $G(q)$. For example, Eq.~(2a) from Ref.~\cite{G4} yields the number of impurities twice as low as the one found in our calculations \cite{rem}.

There is yet another way to check the adequacy of our approach. One can compare the calculated and experimentally found ratios of the resistances in a spin-polarized and spin-unpolarized electron systems. This comparison is especially useful because the effective mass (which is itself strongly density-dependent) cancels out. At $n_\text{s}\gg n_\text{c}$ (\textit{i.e.}, far away from the transition), this ratio is equal to approximately 4 according to both the experiment \cite{kra1} and theory \cite{do}. However, as the density is reduced, this ratio significantly grows, as shown in Fig.~\ref{fig4}. Using the density of impurities obtained above, one can calculate the ratio of the resistances in a spin-polarized and spin-unpolarized systems without using additional adjustable parameters. As seen in the figure, calculations yield not only qualitative but also reasonable quantitative agreement with the experiment. Note that the ratio of resistances depends on temperature, but in the range of electron densities spanned in the figure, this temperature dependence is weak at $T<300$~mK \cite{shashkin2001,met} and can be neglected.

Summarizing the above, one can conclude that the experimental fact --- a spin-polarized electron system localizes at electron densities higher than a spin-unpolarized one --- is a consequence of the exchange-correlation effects. This is in agreement with the conclusions of Ref.~\cite{fl} where the increase of the number of valleys in a strongly-correlated electron system was shown to induce delocalization.

One can expect that the boundary between metal and insulator on an $(n_\text{s},B)$ plane should correspond to the line $n_\text{s}=n_\text{c1}$ when $B\geq B^\text{p}(n_\text{s})$ and to the line $B=B^\text{p}(n_\text{s})$ when $n_\text{s}<n_\text{c1}$. As seen in Figs.~\ref{fig1} and \ref{fig2}, such a behavior is indeed observed in a wide range of parameters except for the immediate vicinity of $n_\text{c}$ and $n_\text{c1}$.

The Anderson transition at densities $n_\text{c}<n_\text{s}<n_\text{c1}$ is of special interest because the localization in this regime comes about in a partially spin-polarized electron system, as can be seen in Fig.~\ref{fig2} where the experimental data \cite{met} corresponding to the complete spin polarization in the insulator are shown. In a certain range of densities between approximately $0.85$ and $1.05\times10^{11}$~cm$^{-2}$, upon the increase of the magnetic field, initially the metal-insulator transition occurs, and only then the full spin polarization is achieved. A similar behavior can also be seen in the data of Refs.~\cite{pu,vitk}.

The exchange interactions between delocalized electrons and interactions between localized electrons are known to yield corrections of the opposite sign to the energy of the ground state. Therefore, in the insulator, a subsequent full spin polarization leads to an increase in the energy by $(n_\text{s}/2)(1-P)\beta e^2\sqrt n_\text{s}/\varepsilon$ (here $P$ is the degree of the spin polarization and $\beta$ is determined by the product of the number of the nearest neighbors and the exchange integral). Full spin polarization will be reached in a magnetic field $B^\text{p}_\text{ins}$ at which the loss in the exchange energy is compensated by the gain in the Zeeman energy, $(n_\text{s}/2)(1-P)g\mu_\text{B}B^\text{p}_\text{ins}$. Thus, one obtains

\begin{equation}
B^\text{p}_\text{ins} = \frac{\beta e^2 \sqrt n_\text{s}}{\varepsilon g \mu_\text{B}}.\label{8}
\end{equation}
In Fig.~\ref{fig2} we show the result of calculations of $B^\text{p}_\text{ins}$ using Eq.~(\ref{8}) with a reasonable value of $\beta=0.06$, which fits the data \cite{met} on the complete spin polarization in the insulator.

In conclusion, behavior of 2D electrons in Si MOSFETs is governed by the competition between interactions and disorder. In the region of metallic conductivity the electron interactions are crucial, leading to such experimentally observed phenomena as the anomalous increase of the effective mass. As the electron density is decreased, the influence of disorder becomes more important and leads to the Anderson transition except for the least disordered samples where Wigner crystallization possibly occurs \cite{bru}. Spin polarization strengthens the disorder effects, which is manifested by the increase in the critical density for the MIT. We have shown that this increase is the result of the exchange and correlation effects. When the metal-insulator transition occurs in a partially spin-polarized electron system, an additional increase in the magnetic field is necessary to achieve the full spin polarization due to the exchange interactions between localized electrons. Determination of the boundary between metal and insulator on the $(n_\text{s},B)$ plane in a partially spin-polarized system requires further work.

We are grateful to I.~S. Burmistrov, E.~V. Deviatov, and V.~S. Khrapai for fruitful discussions. This paper was supported by RFBR Grants No.\ 15-02-03537 and No.\ 16-02-00404, Russian Academy of Sciences, and Russian Ministry of Science. SVK was supported by NSF Grant No.\ 1309337 and BSF Grant No.\ 2012210.


\begin{thebibliography}{90}
\bibitem{rem} The functions $G(q)$ obtained for a single-valley system were mistakenly used in Ref.~\cite{G4}. Therefore, Eq.~(2a) in Ref.~\cite{G4} corresponds to a spin-polarized two-valley system.
\end{thebibliography}
\end{document}